# Design of quasi phase matching crystal based on differential gray wolf algorithm

He Chen1,†, Zihua Zheng1,†, Jinghua Sun1,2,3,*

1.School of Electrical Engineering & Intelligentization, Dongguan University of  Technology, Dongguan, Guangdong 523808, China.

2.Guangdong Provincial Key Laboratory of Advanced Particle Detection Technology, Dongguan, Guangdong, 523000, China.

3.International Aerospace Institute, Dongguan University of Technology, Dongguan, Guangdong 523808, China.

†Authors with equal contribution.

*Corresponding author: sunjh@dgut.edu.cn.

**摘要**：本文聚焦非线性光学技术发展中的关键问题，非周期极化晶体性能优化。该晶体性能依赖晶畴微观分布的精准控制，但其优化属于高难度的高维离散组合"NP 难"问题，传统算法存在收敛慢、易陷局部最优的瓶颈，遗传算法等启发式方法则受 CPU 串行计算限制，效率不足。

为解决上述挑战,本文提出 HWSDA 混合优化算法与 GPU 并行加速技术的融合方案：以差分进化算法（DE）实现全局搜索，搭配灰狼优化算法（GWO）强化局部搜索与收敛速度，二者协同平衡全局与局部优化需求；同时依托 GPU 多核心架构实现线程级并行计算，提升优化效率。

该方案有效突破高维离散空间优化难题，提升了晶畴调控精准度，较传统 CPU 串行计算将准相位匹配设计效率提升数百倍至数千倍,为复杂非线性光学器件设计提供新范式，助力推动量子光学、激光加工等领域相关器件的性能突破与产业化应用。

**关键词**：准相位匹配晶体设计；差分进化算法；灰狼算法；GPU 并行计算；

# 1. 引言

非线性光学技术的突破高度依赖非线性晶体性能的提升，这类晶体通过倍频、和频、差频及光参量振荡等效应，实现光信号的频率调控与谱域拓展，是多领域技术革新的核心基石。在量子光学领域，新加坡国立大学朱迪团队的分层极化铌酸锂纳米波导显著提升了非线性转换效率，为频率调控提供新路径[1]；厦门大学张武虹等通过 Hardy 判据验证高维角位置纠缠态，推动了量子信息传输发展[2]。生物医学成像中，新加坡国立大学刘小刚与厦门大学梁亮亮合作开发的高非线性纳米晶实现亚衍射极限成像[3]；华东师范大学黄坤团队的中红外非线性成像技术突破传统晶体孔径限制，助力生物检测[4]。激光加工方面，Zhang 等人在 LBO 晶体中通过频率混频产生 193nm 的高功率窄线宽固体深紫外激光[5]；Du 等人利用钙钛矿材料因其独特的光学和电学特性系统总结了钙钛矿沉淀和图案化的方法和优势和该综述还为进一步优化和改进钙钛矿飞秒激光制备和加工方法提供了展望[6]。Lei 等人提出了通过高性能计算与第一性原理方法辅助晶体设计的完整流程，尤其适用于深紫外和中红外波段的非线性晶体[7]。

非周期极化晶体作为非线性光学领域的核心功能材料，其倍频效率、带宽覆盖范围及偏振调控精度等关键性能指标，高度依赖于晶畴微观分布的精准调控与优化设计。本质而言，晶畴分布的优化问题隶属于高维离散空间下的组合优化范畴，其解空间维度随晶体长度与畴结构复杂度呈指数级增长，凸显出 NP 难问题的典型特征。

在当前研究中，传统优化算法（如模拟退火算法[8,9]）虽在低维优化问题中展现出一定有效性，但在面对非周期极化晶体的高维离散优化场景时，普遍存在显著瓶颈：一方面，其基于单点迭代的搜索模式难以高效遍历庞大解空间，导致收敛速度迟缓，尤其在晶体畴数超过 $10^3$ 量级时，计算耗时常突破实用阈值；另一方面，算法对初始解的依赖性较强，易因局部能量陷阱陷入次优解，难以保证晶畴分布的全局最优性，直接制约了器件性能的进一步提升。

针对这一困境，部分研究引入遗传算法[19]等启发式智能优化方法[12]，通过种群进化机制拓展搜索广度，在一定程度上缓解了局部最优问题。然而，此类算法的核心优势（如群体多样性维持与全局搜索能力）的充分发挥，高度依赖于并行化计算架构的支撑。当前主流的 CPU 串行执行模式，将种群内个体的适应度评价、交叉变异等操作限制在单线程迭代流程中，不仅导致算法时间复杂度随种群规模呈线性增长，更难以实现多子种群的协同进化与信息交互，极大削弱了启发式算法的全局寻优效能。

更深层次来看，当前研究在非周期极化晶体优化领域，对于 GPU 并行计算架构与物理约束条件的融合探索仍显薄弱。现有方法多聚焦于算法本身的收敛性

改进，而对计算架构的并行化改造与物理模型的约束嵌入缺乏系统性设计，GPU 所具备的众核并行计算能力尚未被充分挖掘，未能通过线程级并行实现种群个体的批量评价与进化操作，导致计算效率提升受限。上述问题的叠加，使得现有优化方法的计算复杂度居高不下，不仅增加了非周期极化晶体的设计周期与制造成本。

上述挑战的核心在于如何在保证优化精度的前提下，突破算法效率与计算架构的双重限制。其中，差分进化算法（DE）因其在连续空间优化中展现的独特优势受到广泛关注：其核心优势在于原理简洁、全局搜索能力突出，尽管原生设计并非针对组合优化场景，但可通过数值映射关系适配非周期极化晶体的离散畴结构优化需求；更重要的是，DE 的种群并行性与 GPU 的线程级并行特性高度契合，通过 GPU 加速可显著压缩运行时间，为大规模畴结构优化提供算力支撑。不过，DE 在处理非周期极化晶体这类兼具连续参数优化与离散结构设计特征的复杂问题时，仍存在后期易陷入局部最优的缺陷，且对局部解的精细挖掘能力有限[13]。

与之形成互补的是灰狼优化算法（GWO），其模拟狼群狩猎的寻优机制，具备较强的局部搜索能力与快速收敛特性，能够精准挖掘局部最优解，恰好可弥补 DE 在局部开发阶段的不足。但是 GWO 算法同样存在问题，在求解高维函数容易陷入局部最优[14]，也缺乏陷入局部最优后比较难以摆脱当前位置的策略[15]，因此，构建 DE 与 GWO 的混合狼群差分算法（Hybrid Wolf Swarm Differential Algorithm，HWSDA），并深度融合 GPU 并行加速技术，有望实现全局勘探和局部开发的协同优化，既通过 DE 的种群进化机制拓展解空间覆盖范围，又借助 GWO 的精细搜索能力提升解的质量，同时依托 GPU 的众核计算架构大幅缩短优化周期。这一技术路径为非周期极化晶体的晶畴分布优化提供了兼具高效性与精准性的解决方案。

## 2. 相关技术与基础知识

### 2.1 非周期极化晶体优化的物理模型

这里将以耦合三倍频为起点推导一下后面 DGWO 算法需要寻优的目标函数。在光学系统中，耦合三倍频包括了先倍频（SHG）后和频（SFG）的一个过程。忽略色散项，在平面波近似下，其耦合波方程为[18]：

$$\frac{dE_1}{dz} = \frac{i\omega_1 d_{33}}{n_1 c} d(z) E_1^* E_2 \cdot exp(i\Delta k_1 z) + \frac{i\omega_1 d_{33}}{n_1 c} d(z) E_2^* E_3 \cdot exp(i\Delta k_2 z)$$

$$\frac{dE_2}{dz} = \frac{i\omega_2 d_{33}}{2n_2 c} d(z) E_1^2 \cdot exp(-i\Delta k_1 z) + \frac{i\omega_2 d_{33}}{n_2 c} d(z) E_1^* E_3$$
$$\cdot exp(i\Delta k_2 z)$$

$$\frac{dE_3}{dz} = \frac{i\omega_3 d_{33}}{n_3 c} d(z) E_1 E_2 \cdot exp(-i\Delta k_2 z)$$

其中 i 为虚数单位，$\Delta k_1$ 和 $\Delta k_2$ 分别为 SHG 和 SFG 的相位失配量，式中 $d_{33}$ 为铌酸锂（PPLN）33 方向的非线性系数，c 为真空中的光速，$\omega_1$、$\omega_2$、$\omega_3$ 分别为基频光、倍频光和三倍频光的频率，$E_1$、$E_2$、$E_3$ 分别为基频光、倍频光和三倍频的振幅。d(z)为对应晶畴的极化方向，目前设定+1 为向上，-1 为向下。假设在进行耦合三倍频的过程中，基频光的能量不发生损耗，则存在 $E_1$ 等于一个常数。同时假设非线性光学转换不强，则可能存在 $E_3$ 远小于 $E_2$，则：

$$\frac{dE_2}{dz} = \frac{i\omega_2 d_{33}}{2n_2 c} d(z) E_1^2 \cdot exp(-i\Delta k_1 z) + 0,$$

假设晶体总长度为 L，对上式进行积分可得：

$$\int_0^L dE_2 = \frac{i\omega_2 d_{33}}{2n_2 c} E_1^2 \int_0^L d(z) \cdot e^{-i\Delta k_1 z} dz$$

将等式左边求解可得：

$$E_2(L) = \frac{i\omega_2 d_{33}}{2n_2 c} E_1^2 \int_0^L d(z) \cdot e^{-i\Delta k_1 z} dz$$

将其带入第三个式子可得：

$$\int_0^L dE_3 = \frac{i\omega_3 d_{33}}{n_3 c} E_1 \left( \frac{i\omega_2 d_{33}}{2n_2 c} E_1^2 \int_0^z d(x) \cdot e^{-i\Delta k_1 x} dx \right) \int_0^L d(z) \cdot e^{-i\Delta k_2 z} dz$$

整理可得：

$$E_3(L) = -\frac{\omega_2 \omega_3 d_{33}^2}{2n_2 n_3 c^2} E_1^3 \int_0^L d(z) e^{-i\Delta k_2 z} \left( \int_0^z d(x) e^{-i\Delta k_1 x} dx \right) dz$$

可以看出，耦合三倍频的输出光的强度正比于

$$E_3(L) \propto d_{eff} = \int_0^L d(z) e^{-i\Delta k_2 z} \left( \int_0^z d(x) e^{-i\Delta k_1 x} dx \right) dz$$

其中，i 为虚数单位，d(x)为晶畴第 x 个晶畴的方向取向(+1/−1)，$\Delta k_1$ 为 SHG 的相位失配量，$\Delta k_2$ 为 SFG 的相位失配量。由上述公式可以看出，想要输出光的效率越高与 d(x)的排列存在着很重要的关系，因此我们需要找出合适的 d(x)，使得 $d_{eff}$ 最大。

## 2.2 基础优化算法原理

### 2.2.1 基本差分进化算法

差分进化算法（Differential Evolution，DE）是一种新兴的进化计算技术。在 1997 年，由 Storn 等人提出用来解决切比雪夫多项式问题[14]，后来发现它也是解决复杂优化问题的有效技术。差分进化算法和遗传算法很相似，也是一种基于群体智能理论的优化算法，通过群体内个体间的合作与竞争而产生的全局搜索策略，采用实数编码、基于差分的简单变异操作和"一对一"的竞争生存策略，降低了进化计算操作的复杂性。

DE 算法凭借种群记忆机制与无模型优化特性，可动态感知搜索状态并自适应调整勘探策略，在未知复杂优化空间中展现出优异的全局收敛能力与鲁棒性，无需依赖问题的梯度、凸性等先验特征，成为求解非凸、多峰、高维离散/连续优化问题的核心工具（尤其在传统数学规划方法失效的场景中）。与遗传算法（Genetic Algorithm, GA）等进化算法类似，DE 通过种群级候选解的迭代演化 逼近最优解，但其核心操作（变异、交叉）的设计更聚焦差分信息的利用，形成了独特的进化范式，具体包含以下关键步骤：

1. 种群初始化：初始化阶段需要为每个种群（规模为 NP）的每个个体（维度为 D）分配初始值，以保证解的多样性与解空间的覆盖性，往往采用均匀分布采样，数学描述为：

$$x_{i,j} = U(X_{min,j}, X_{max,j}), i = 1,2,\ldots,NP; j = 1,2,\ldots,D$$

其中，$X_{min,j}$ 和 $X_{max,j}$ 分别为第 j 维度的上下界。

2. 变异操作（Mutation）：通过种群内个体的差分信息构造新候选解，核心公式（以经典 rand/1 策略为例）为：

$$v_i^{t+1} = x_{r_1}^t + F \cdot (x_{r_2}^t - x_{r_3}^t)$$

其中，r1,r2,r3 为所有种群的索引，要求 r1,r2 和 r3 互不相同，且与 i 也不能相同，所以这要求种群数量 $NP \geq 4$,变异算子 F∈[0,2]是一个实常数因数，它控制

偏差变量的缩放，决定全局勘探步长。

3. 交叉操作（Crossover）：将变异向量的参数与另外预先确定的目标向量参数按照一定的规则混合来产生试验向量。

$$u_{ji}^{t+1} = \begin{cases} u_{ji}^t, & if\ randb(j) \leq CR\ OR\ j = rnbr(i) \\ x_{ji}^t, & x \neq 0\ and\ j \neq rnbr(i) \end{cases}$$

其中，randb(j)表示产生[0,1]之间的随机数第 j 个估计值。rnbr(i)代表着在维度 D 中随机取一个。CR 表示交叉算子，其取值范围为[0,1]。通过维度级随机混合，交叉操作在变异引入的全局差分信息基础上，进一步丰富种群的局部多样性，防止算法过早陷入单一进化路径。

4. 选择操作（Choose）：选择操作采用贪婪策略，仅当试验向量的适应度更优时，才替换目标向量，数学描述（以最小化问题为例）为：

$$X_i^{new} = \begin{cases} v_i, & if\ f(v_i) < f(x_i) \\ u_i, & else \end{cases}$$

其中，f(x)为适应度函数。该机制确保每代种群的适应度单调不减（最小化场景），持续积累更优解的进化压力，同时保留较优的历史个体，避免有效信息的丢失。

DE 算法的种群演化机制，天然孕育了个体级计算解耦的并行范式，天然适配 GPU 并行加速，每一代种群的所有个体均作为目标向量，通过变异、交叉操作生成等规模的竞争个体，严格维持种群规模的恒定迭代；进化过程中，逐代对向量的适应度进行评估，驱动优化目标（如最小化问题）向全局最优解收敛。这种基于随机偏差扰动的新个体生成策略，不仅赋予算法优良的收敛特性，更构建了核心操作（初始化、变异、交叉、选择及适应度计算）的强并行性基础。

### 2.2.2 基本灰狼算法

灰狼优化算法(Grey Wolf Optimizer,GWO)[15]是 2014 年提出的一种群体智能算法，GWO 算法模拟自然界中灰狼种群等级机制和捕猎行为。通过 4 种类型的灰狼来模拟社会等级，通过狼群跟踪、包围、追捕、攻击猎物等过程来模拟狼的捕猎行为,实现优化搜索目的。GWO 算法具有原理简单、并行性、易于实现,需调整的参数少且不需要问题的梯度信息,有较强的全局搜索能力等特点。

在 GWO 算法中,首先是在搜索空间中随机产生灰狼族群,为构建灰狼的社会等级制度模型,将种群中适应度值最优的解、次优的解和第三优的解分别看作α

狼、β狼和γ狼,而剩余的解被视为ω狼。在不断的迭代搜索靠最优的三个头狼来预测最优解。在搜索过程中，灰狼种群捕食的更新公式如下

$$d = |C \cdot X_p(t) - X(t)|$$

$$X(t+1) = X_p(t) - A \cdot d$$

其中，d 代表着距离，A、C 表示系数向量，t 为迭代次数，$X_p$ 表示当前猎物的位置向量，X 表示其中灰狼的位置。系数向量 A、C 的表达式为

$$A = 2 \cdot a \cdot r_1 - a$$

$$C = 2 \cdot r_2$$

其中，$r_1$ 和 $r_2$ 是 [0,1] 内的随机数，系数 a 随着迭代次数从 2 到 0 线性递减。在寻优的过程中，将α狼、β狼和γ狼视最优解，将其余的狼群视为ω狼，利用前面的位置公式对ω狼进行更新。

$$d_\alpha = |C_\alpha \cdot X_\alpha - X|$$

$$d_\beta = |C_\beta \cdot X_\beta - X|$$

$$d_\gamma = |C_\gamma \cdot X_\gamma - X|$$

其中，$d_\alpha$、$d_\beta$ 和 $d_\gamma$ 代表着当前狼与前三只头狼的近似距离，通过其距离计算当前解的最终位置公式如下

$$X_1 = X_\alpha - A_1 \cdot d_\alpha$$

$$X_2 = X_\beta - A_2 \cdot d_\alpha$$

$$X_3 = X_\gamma - A_3 \cdot d_\alpha$$

$$X(t+1) = \frac{\omega_1 X_1 + \omega_2 X_2 + \omega_3 X_3}{3}$$

其中，$A_1$、$A_2$ 和 $A_3$ 是随机产生系数向量，$X_\alpha$、$X_\beta$ 和 $X_\gamma$ 分别为α狼、β狼和γ狼的位置，t 代表当前的迭代次数。其中灰狼种群中 3 个解的位置权重为

$$\omega_1 = \frac{|X_1|}{|X_1| + |X_2| + |X_3|}$$

$$\omega_2 = \frac{|X_2|}{|X_1| + |X_2| + |X_3|}$$

$$\omega_3 = \frac{|X_3|}{|X_1| + |X_2| + |X_3|}$$

通过当前表现最好的 3 种狼所处的位置来锁定狼群所捕对象所处的区域位置，即在 GWO 算法中以当前 3 个最好解的位置来综合估计预测全局最优解的位置，并通过其权值动态调整不同级别的灰狼对最优值的影响。因此与其他智能算法只在单个解位置带领下寻最优解相比，在很大程度上有效地降低了算法在搜索前期陷入局部极值的概率。

GWO 作为一种群智能优化算法，其核心操作的强并行性与 GPU 的硬件特性同样也高度适配：算法的初始化、适应度计算、个体位置更新等核心步骤中，种群内每个灰狼个体的操作（如位置初始化、基于前三只头狼的个体位置调整、适应度评估等）相互独立，几乎无跨个体的依赖关系，可并行执行。GPU 拥有大量并行线程，能同时调度多个线程处理不同个体的计算，尤其在大规模种群或高维问题中，可显著加速耗时的适应度计算和位置更新环节，同时算法中个体间信息交互少、无需频繁全局同步，能充分契合 GPU 的内存架构和并行计算模式，从而高效发挥硬件优势。

## 2.3 GPU 并行计算基础

CUDA 架构通过异构计算范式构建了主从式协同计算模型，形成了控制流与数据流分离的双层计算模型。在这个模型中，主机端（CPU）作为控制核心负责执行逻辑密集型任务，包括算法参数的初始化、迭代流程控制、种群统计分析等串行操作，而设备端（GPU）作为计算引擎通过单指令多线程（SIMT）架构执行数据密集型任务，实现大规划并行加速。这种分工模式充分发挥了 CPU 的复杂逻辑处理能力与 GPU 的高度并行计算能力，形成优势互补的协同计算体系。

CUDA 核函数的执行往往采用三级层次化线程抽象，构建了从宏观到微观的并行计算结构：

1. 网格（Grid）：作为最高层抽象，代表一次核函数调用所创建的完整线程集合，映射算法层面的并行任务单元。

2. 线程块（Block）：作为中间层抽象，是共享内存分配与线程同步的基本单位，通过线程协作机制实现细粒度数据交互（如领域个体信息交换）。

3. 线程（Thread）:作为最底层抽象，是指令执行的最小单元，负责处理单条数据记录或计算子任务（如单个个体的变异操作）。

4. 这种层次化设计通过线程索引系统实现对计算资源的精确寻址，结合线程束（Wrap）机制（32 个线程为一组同步执行），既保证了并行计算的灵活性，

又优化了硬件执行效率。

CUDA 的内存系统采用层次化存储模型，构建了不同访问速度与容量的存储层次：

1. 寄存器（Registers）:每个线程私有存储，提供最快的访问速度，用于存储线程局部变量。

2. 共享内存（Shared Memory）:线程块内共享的片上缓存，具备高带宽（数百 GB/s）与低延迟特性，通过数据分块与缓存优化策略支持线程间高效协作；

3. 全局内存（Global Memory）:设备端主存，提供大容量存储，但访问延迟较高（数百周期），需通过合并访问、内存对齐等技术提升带宽利用率；

4. 常量内存（Constant Memory）与纹理内存（Texture Memory）：针对读多写少场景优化的特殊内存，通过硬件缓存机制加速数据读取。

在进化算法并行实现中，本文采用内存分层优化策略：将频繁访问的种群核心参数（如个体适应度值）存储于共享内存，利用线程块内的数据复用特性减少全局内存访问；通过合并访存模式组织线程对全局内存的读写操作，使连续线程访问连续内存地址，充分利用 DRAM 的突发传输特性。针对算法中的只读参数（如非线性光学系数），利用常量内存的广播机制实现高效数据分发。

针对进化算法的并行加速需求，本文将适应度计算、变异操作、交叉操作等核心步骤，按种群个体维度分解为线程任务：单个线程独立处理单一个体的计算或更新，线程块内通过共享内存实现局部数据的快速同步，全局内存保障跨块数据的持久化存储与主机交互。该策略充分挖掘 GPU 的大规模线程并行能力，在高维、大种群规模场景下，显著提升算法迭代效率，突破 CPU 串行计算的性能瓶颈。

## 3. 基于 GPU 加速的混合狼群差分算法设计

### 3.1 改进差分灰狼算法

#### 3.1.1 双机制的自适应搜索策略

这里提出一种基于种群动态特性的变异因子 F 自适应调整策略，其核心逻辑通过整合种群多样性指标（标准差、适应度范围）、收敛状态（收敛率）及迭代进度，实现 F 在[0.01,0.1]区间内的动态优化，以平衡算法的探索能力与开发精度。具体而言：当种群多样性较低（标准差小于某个阈值或收敛率低于 0.1）时，通过增强探索因子适度提升 F，避免算法陷入局部最优。当种群多样性过高或适应度范围较小时（标准差大于某个阈值或适应度范围小于 $10^3$），则通过开发因子降低 F，聚焦优质解的精细化搜索。对于中间状态，采用基于迭代进度的余弦衰减函数使 F（公式如下）自然下降，同时叠加二次衰减系数（随迭代进度从 1.0

降至 0.8），确保后期 F 稳定收敛至较低水平。

$$F = \frac{\cos(\frac{\pi g}{2G}) + 1}{2}$$

其中，g 为当前迭代次数，G 为总迭代次数。

这一设计的核心目的在于：变异因子 F 直接影响种群中解的更新幅度，在高维优化问题（如非线性晶体晶畴结构优化）中，F 过大易导致过多决策变量（晶畴）发生状态翻转，破坏相位匹配条件而降低优化效率；F 过小则可能限制算法的探索能力，难以跳出局部最优。通过耦合种群实时状态（多样性、收敛程度）与迭代阶段的自适应调整，该函数可在优化前期维持适度探索以覆盖解空间，后期通过降低 F 减少不必要的决策变量扰动，最终在满足物理约束（如避免过量晶畴翻转）的同时，提升算法的收敛精度与稳定性。

### 3.1.2 改进离散灰狼算法

该算法创新性地借鉴自然界狼群社会中头狼层级动态更替的生态特征，将传统灰狼算法的三头头狼架构扩展为四头头狼协同引导机制。这一改进不仅契合狼群生态中备选头狼竞争上位的自然规律，更通过增加头狼群体基数拓展了算法前期的搜索覆盖范围。与传统 GWO 相比，四头头狼的设置使得离散状态空间的探索样本更为丰富，能够在迭代初期捕捉更多潜在的优质解区域，有效缓解了传统算法因引导个体不足导致的早期搜索偏向性问题。

算法专门针对离散特性设计核心机制：将连续位置值直接映射为+1/-1 离散状态，替代传统 GWO 的连续位置更新；通过统计四头头狼的状态分布计算状态转移概率（而非传统 GWO 中基于距离的连续位置调整），基于迭代进度动态调整搜索策略：迭代前期以高概率随机扰动促进全局探索（比传统 GWO 的线性参数衰减更灵活），基于四头头狼的状态分布进行概率转移，并通过离散度参数控制扰动强度。迭代后期倾向优势状态并减少扰动以聚焦开发（弥补传统 GWO 在离散域易过度震荡的缺陷），优先选择四头头狼中占优的状态，仅保留小概率翻转机制。
同时引入社会学习机制，允许个体以随迭代递减的社会学习率直接模仿四头头狼中的随机个体状态，强化优质离散结构的保留，传统 GWO 缺乏此类针对性机制。通过离散度因子、社会学习率和扰动率等参数的协同变化，改进算法解决了传统 GWO 在离散域中连续更新与离散状态不匹配和探索开发平衡僵化的问题，结合 CUDA 并行框架，大幅提升了高维离散问题的求解效率与精度。

### 3.2 算法整体步骤

综合上述改进策略描述，给出本文所提出的改进算法的具体步骤：

**步骤** 1 设置种群规模 NP，最大迭代次数 G，对参数 A、F 进行初始化等

**步骤2** 利用变异操作、交叉操作和选择操作得到所有个体的适应度值

**步骤3** 直接在 GPU 上进行并行约归操作，最所有个体的适应度函数值进行排序，选出最好的前三只头狼，记为 $X_\alpha$、$X_\beta$ 和 $X_\gamma$。

**步骤4** 利用灰狼算法中的公式更新种群中的其他灰狼个体的位置。

**步骤5** 再次进行差分算法中的选择操作，用于保存最优狼群。

**步骤6** 对 F、CR 等参数进行更新。

**步骤7** 判断算法是否达到最大迭代次数 G，若以达到最大迭代次数则结束计算，输出最优结果。否则，重复执行**步骤2**到**步骤6**。

## 4. GPU 加速算法的必要性分析

在处理此类高维复杂优化问题时，由于解空间维度的激增与目标函数的非线性耦合，算法的计算开销常随问题规模呈指数级增长，极大制约了其在工程实践中的应用效能。在此背景下，GPU 的并行加速技术为突破这一性能瓶颈提供了关键支撑，其硬件架构与算法内在并行特性的深度适配，成为提升计算效率的核心路径。

具体而言，差分进化算法（DE）的变异、交叉与选择操作，以及灰狼优化算法（GWO）的社会层级引导（如 α、β、δ 狼的层级信息传递）与维度更新机制，均蕴含天然的并行计算潜力。在 DE 中，个体的变异向量生成仅依赖种群中随机选取的少量个体信息，交叉与选择操作亦以个体为独立单元执行，各体间无强依赖关系；在 GWO 中，普通个体（β、δ 狼）的位置更新虽需参考头狼（α 狼）的全局最优位置，但每个个体的维度级调整可独立完成，无需线程间的同步等待。这种无依赖并行数据流特性，使得两类算法的核心操作可分解为大量同构的子任务，与 GPU 的单指令多线程（SIMT）架构高度契合，算法中统一的操作规则可通过单条指令驱动大规模线程阵列同步执行，显著提升计算资源的利用率与指令吞吐量。

现代 GPU 作为面向数据密集型计算的专用硬件架构，其核心优势体现在大规模并行计算单元与高带宽内存系统的协同设计：主流 GPU 通常集成数千个流处理器核心，并通过多线程束调度机制实现数万线程的并发执行，配合数百 GB/s 级别的内存带宽，可高效支撑高维优化问题中密集的数据读写与计算需求。这种硬件特性与混合狼群差分算法的种群级并行特性形成深度契合。GPU 的单指令多线程（SIMT）架构通过统一指令流驱动多个线程束同步执行，完美适配差分灰狼算法中种群内个体的同构操作，例如：在种群规模为 10000 的优化场景中，每个个体的适应度计算、变异更新等操作可被分配至独立线程，由同一指令流控

制执行，避免了 CPU 串行模式下的线程切换开销与资源闲置。对于 10000 维等高维问题，GPU 可进一步实现维度级并行：将每个维度的计算任务映射至独立线程，通过线程束内的并行调度完成向量级运算，而 CPU 受限于核心数量与内存带宽，往往只能通过循环迭代处理数十个维度，导致计算效率存在量级差距。

在晶体畴工程等涉及数万维度的优化场景中，传统 CPU 串行实现可能耗时数小时甚至数天，而 GPU 加速可将计算时间压缩至分钟级，使实时优化和大规模参数扫描成为可能。

## 5. 实验与结果分析

本文所有实验均部署于 Windows 操作系统环境下。在硬件配置方面，中央处理器（CPU）选用英特尔酷睿 i5-12400F，作为计算机的运算核心，其具备 6 核心 12 线程的架构设计，能够为实验中涉及的串行计算任务（如数据预处理、结果后处理等）提供稳定的算力支撑，保障基础程序逻辑的高效执行。

图形处理器（GPU）采用英伟达 GeForce RTX 4070，其配备 12GB 显存，显存作为 GPU 临时存储数据的空间，12GB 的容量可满足大规模数据（如算法种群数据等）在 GPU 端的缓存需求，减少因数据频繁在 CPU 与 GPU 之间传输造成的性能损耗。同时拥有 5888 个 CUDA 核心，CUDA 核心是英伟达 GPU 实现并行计算的基本单元，数量众多的核心可同时并行处理多个计算任务（如本文算法中适应度计算、种群更新等并行化操作），大幅提升计算密集型任务的执行效率；显存类型为 GDDR6/GDDR6X，这类显存凭借高带宽、低延迟的特性，能够高效读写数据，进一步优化 GPU 处理数据的速率，为实验中基于 GPU 加速的算法提供了有力的硬件加速基础，助力算法在复杂解空间中快速探索与迭代。下表展示了种群均设置为 5000 个的情况下对于不同维度，相同算法在 GPU 与 CPU 的运行时间对比。

表 1 HWSDA 在 GPU 与 CPU 运行时间对比表

| Domain dimension | HWSDA（GPU） | HWSDA（CPU） | Acceleration ratio |
|---|---|---|---|
| 660 | 2.0s | 3197.4s | 1598.7 |
| 1320 | 4.7s | 7016.82s | 1492.9 |
| 3300 | 11.8s | 21550.45s | 1826.3 |
| 6600 | 22.5s | 43713.61s | 1942.8 |
| 13200 | 45.3s | 70325.64s | 1552.4 |

本文还开展了线程块配置对算法 GPU 加速性能的影响实验，并对不同配置下的运行时间进行了系统性对比分析。实验过程中，保持种群规模（NP）、变异因子（F）、交叉概率（CR）等核心算法参数与前文实验设置一致，仅通过调整线程块维度配置作为变量，以孤立分析其对算法运行时间的影响规律。鉴于所采用显卡硬件特性，在 GPU 架构中线程束（Warp）的硬件设计固定为 32 个并行

线程，因此，本实验选取线程块规模为 32 的整数倍（即 32×n，其中 n 为正整数）作为测试变量，以探究在硬件架构约束下线程块配置与运行效率的关联机制。

具体实验场景设定为：晶体长度 660μm、晶畴厚度 0.1μm，泵浦光波长 1404nm 进行耦合三倍频。为降低随机因素对实验结果的干扰，上述测试条件下的每种线程块配置均独立重复运行 30 次，以获取具有统计意义的时间基准数据，实验结果将为该算法在同类硬件平台上的最优线程配置提供量化依据。

表 2 不同线程数对于运行时间的影响

| 线程数量 | 平均时间 |
| --- | --- |
| 32 | 21.3s |
| 64 | 21.7s |
| 128 | 21.2s |
| 256 | 23.9s |
| 512 | 33.1s |
| 1024 | 59.8s |

通过对实验数据的深入分析可以发现，在种群规模为 5000 的实验场景中，GPU 线程数量与算法运行速度并非呈现简单的线性正相关关系。在 GPU 架构体系里，存在多个流式多处理器（Streaming Multiprocessors，SM），其作为 GPU 并行计算的核心单元，负责线程的调度与执行。GPU 会依据自身硬件资源分配机制，将所有线程动态分配到各个 SM 中。然而，当某个线程束对共享内存的占用需求过大时，会引发资源竞争，对其他线程的内存访问造成干扰，出现资源抢占现象，进而导致整体计算效率下降。这一现象深刻揭示了 GPU 并行计算中硬件资源分配与任务特性适配的复杂性。线程数量的选择需综合考量任务的计算复杂度、内存访问模式、数据依赖关系等多维度因素。对于本算法而言，经过对不同线程数量配置下的运行效率、资源利用率等指标进行系统性测试与分析，发现在线程数量为 128 时，能够实现硬件资源（如 SM 利用率、共享内存占用、寄存器使用等）与算法计算需求的最优匹配。此时，线程束的内存访问冲突得以有效规避，SM 的并行计算能力得以充分发挥，算法在 GPU 加速环境下的运行效率达到较优状态。故而，本算法最终选定 128 个线程作为核心配置，以保障在种群规模为 5000 的实验及应用场景中，算法具备高效、稳定的并行计算性能。

为验证算法性能及 GPU 运行效率，实验选取非线性晶体有效非线性系数求解作为目标函数，采用独立运行 30 次的统计分析方法：通过均值刻画在给定迭代次数下的收敛精度，以运行时间、标准差衡量算法效率与稳定性。参数设置方面，综合非线性晶体的物理特性与高维优化需求，进行如下定制化设计。

利用 GPU 的大规模并行架构，5000 个种群可通过线程级并行高效计算，避免 CPU 串行的性能瓶颈，同时大种群能覆盖更广泛的初始解分布，降低算法陷

入局部最优的概率，尤其在高维（如1000维晶畴）场景中，更需充足的种群多样性支撑全局探索，最大迭代次数设为1000次，通过前期预实验验证：该迭代次数既能保证算法充分收敛（跟踪适应度曲线的平稳段判断收敛状态），又能控制实验时间成本，实现收敛深度与计算效率的平衡。

传统差分进化（DE）与灰狼优化（GWO）中，F和a通常取[0,2]的范围，但针对非线性晶体的高维优化场景（如1000维晶畴分布），大参数会引发晶畴翻转数爆炸，当F=a=0.1时，每次迭代平均仅10到20个晶畴发生状态翻转（+1/-1切换），而当参数增至0.5时，翻转数骤升至200到300个。从晶体物理机制分析这方面来看,晶畴翻转会破坏相位匹配条件(非线性光学相互作用的核心前提)畴壁散射、相位累积偏差的叠加效应，将导致倍频、和频效率因相位失配急剧下降。因此，本文对F和a采用初始约束与迭代衰减的策略，将F = a = 0.1，从低参数起步，避免早期大量晶畴翻转破坏解的可行性。随着迭代推进，参数0.1将会逐步下降至0.01，既保留前期对解空间的适度探索，又可以在后期通过低翻转率稳定优质解的相位结构。交叉概率CR初始设置为0.9，这样设置是为了算法初期通过搞概率交叉操作促进每个种群内解的基因交换，维持解的多样性。

同时，为了进一步验证本文提到的混合智能算法的有效性，将其与DE算法与GWO算法进行比较。下面提到的改进DGWO算法，DE算法与GWO算法均在GPU上执行。

表 3 The results of 30 separate experiments conducted with a crystal length of 660um a domain size of 1um, and a pump wavelength of 1404nm

|  | average | max | min | std | time | average $d_{eff}$ |
|---|---|---|---|---|---|---|
| HWSDA | 26019.5 | 26098 | 25963 | 41.1 | 2.0s | 0.25374 |
| DE | 25698.9 | 25906 | 25168 | 194.0 | 1.7s | 0.25064 |
| GWO | 727.1 | 875 | 606 | 73.4 | 1.2s | 0.00708 |

表 4 The results of 30 separate experiments conducted with a crystal length of 660um a domain size of 0.5um, and a pump wavelength of 1404nm

|  | average | max | min | std | time | average $d_{eff}$ |
|---|---|---|---|---|---|---|
| HWSDA | 27455.2 | 27519 | 27369 | 46.4 | 4.8s | 0.25381 |
| DE | 24805.5 | 26325 | 22306 | 1092.4 | 3.1s | 0.22931 |
| GWO | 368.8 | 412 | 332 | 28.5 | 2s | 0.00295 |

表 5 The results of 30 separate experiments conducted with a crystal length of 660um a domain size of 0.2um, and a pump wavelength of 1404nm

|  | average | max | min | std | time | average $d_{eff}$ |
|---|---|---|---|---|---|---|
| HWSDA | 27847.3 | 27930 | 27734 | 55.3 | 12.1s | 0.25364 |
| DE | 19952.5 | 22433 | 17510 | 1938.2 | 8.1s | 0.18174 |
| GWO | 162.1 | 190 | 140 | 16.5 | 4.9s | 0.00147 |

表 6 晶体长度 660um，晶畴 0.1um，泵浦光波长为 1420nm 的单独执行 30 次实验结果

|  | 平均值 | 最大值 | 最小值 | 标准差 | 平均时间 | 平均$d_{eff}$ |
|---|---|---|---|---|---|---|
| HWSDA | 25828.5 | 25915 | 25750 | 46.7 | 22.8s | 0.2507 |
| DE | 12882.5 | 15613 | 9778 | 1919.6 | 16.8s | 0.12504 |
| GWO | 80.5 | 87 | 72 | 5.28 | 8.6s | 9.2281E-05 |

表 7 晶体长度 500um，晶畴 0.1um，泵浦光波长为 2080nm 的单独执行 30 次实验结果

|  | 平均值 | 最大值 | 最小值 | 标准差 | 平均时间 | 平均 $d_{eff}$ |
|---|---|---|---|---|---|---|
| HWSDA | 2376.6 | 2392 | 2338 | 11.3 | 17.5s | 0.24203 |
| DE | 1588.3 | 1832 | 1165 | 199.2 | 13.4s | 0.16178 |
| GWO | 12.4 | 13 | 12 | 0.48 | 6.6s | 0.00130 |

从实验结果能够清晰观测到，随着解空间维度的提升，DE 算法的固有缺陷逐渐凸显。在高维环境下，DE 算法因自身搜索机制的特性，易过早收敛于局部最优区域，难以对解空间进行充分且全面的探索，导致其在复杂高维场景中寻优能力受限，陷入局部极值的概率显著增加。

与此同时，GWO 算法在面对高维优化任务时，也暴露出明显不足。尽管其模拟狼群狩猎的寻优模式具备一定的全局搜索潜力，但在高维解空间的复杂拓扑结构下，单纯依靠该算法的原始搜索策略，难以精准定位到全局最优解的邻域范围，搜索效率与精度均难以满足高维优化需求。

然而，通过将 DE 算法与 GWO 算法进行有机融合所形成的改进算法，展现出了更为优异的搜索性能。这种融合策略巧妙地整合了两种算法的优势：一方面，借助 DE 算法的差分变异操作，为种群引入更丰富的多样性，有效避免算法陷入

局部停滞；另一方面，利用 GWO 算法模拟狼群协作的全局搜索机制，增强算法对全局最优区域的勘探能力。二者协同作用，使改进算法在高维解空间中，既拥有了跳出局部最优的能力，又具备了捕捉全局最优的能力，显著提升了搜索性能，为复杂高维优化难题提供了更具潜力的求解路径。

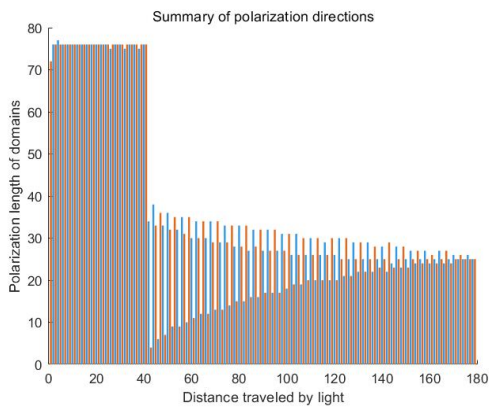
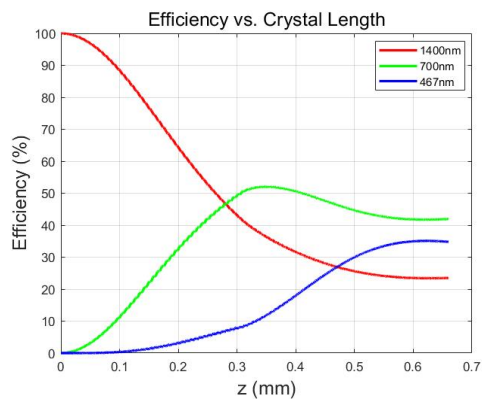
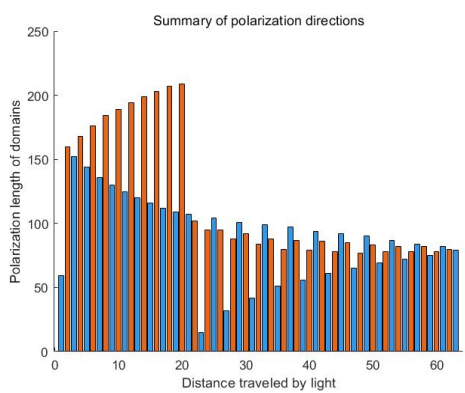
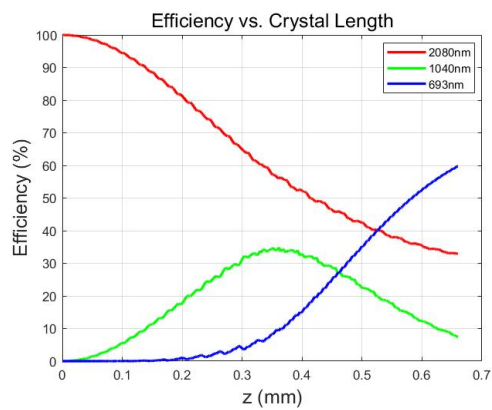
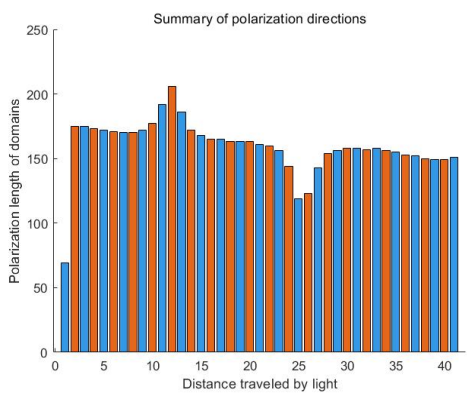
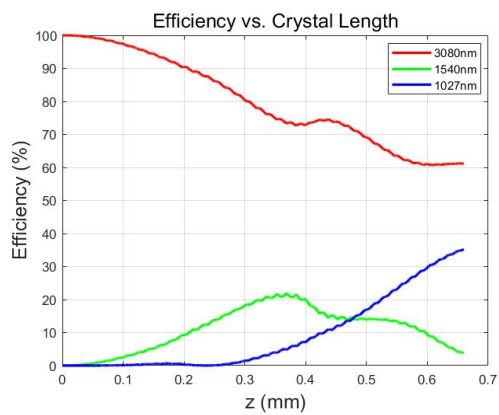

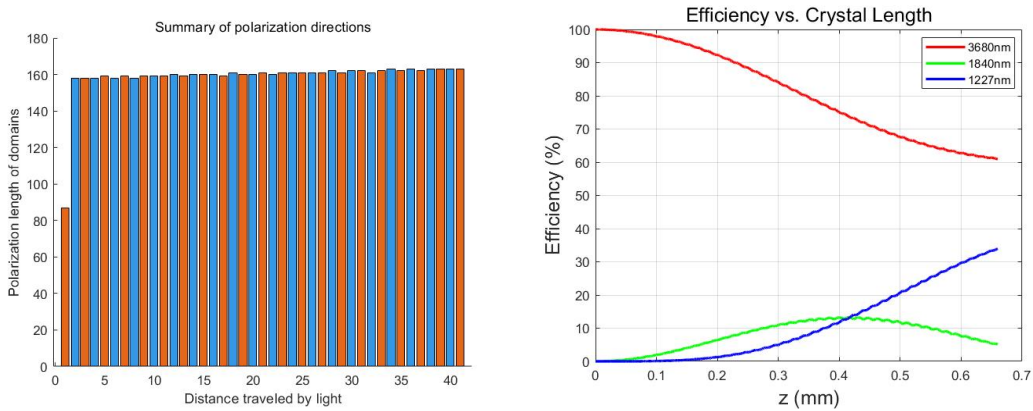

图 6 图中分别展示了 660 微米和波长分别为 1400nm(a)、2080nm(b)、3080nm(c)、3680nm(d)四种波长设计后的晶畴分布以及经过各波长经过晶体的效率变化曲线图

同时本文还尝试利用算法在多波长二倍频中应用此算法来设计非线性晶体，由于在多光二倍频中会存在多个有效非线性系数的组合，因此目标函数改为

$$f = \sum_{i=1}^{n} |G_0 - G_i| + \beta (G_{max} - G_{min})$$

其中，$G_0$为一个比较大的数，$G_i$为设置多波长中的某一个非线性系数值，$\beta$为可调节参数，$G_{max}$为多波长中最大的非线性系数，$G_{min}$同理。

在面向多波长二次谐波产生的非周期性光学超晶格设计与验证实验体系构建中，波长参数的选取参考了文献[9]的波长选取，即选定 972 nm、1082 nm、1283 nm、1364nm 和 1568nm 作为目标入射基波波长。需要说明的是，该文献中所采用的晶体为 $LiTaO_3$，而本实验为选用 PPLN 作为主要晶体。由于不同晶体在介电特性、非线性光学系数、折射率色散等关键物理参数上存在固有差异，依据非线性光学过程中准相位匹配[10, 11]的基本原理，晶体物理属性的不同会直接影响基波与二次谐波在传播过程中的相位匹配条件，进而导致实验结果与文献的结果存在潜在差异。实验采用物理长度为 9mm 的 PPLN 晶体，其周期性反转的晶畴单元厚度设定为 3um，维度为 3000，这一参数选择既考虑了基波与谐波在晶体中的相干长度匹配需求，也兼顾了实际微加工工艺的可行性。测试环境严格控制在 25 摄氏度，以确保晶体折射率的稳定性。在该范围内我们以 0.5 步长进行采样，最后得到如下晶畴灰度图（只截取其中一部分），在预设的五个基波波长处均出现了显著的二次谐波响应峰，且峰值波长与目标值的偏差控制在极小范围内，满足多波长并行 SHG 对 QPM 条件的严苛要求。这一结果表明，通过优化算法构建的 AOS 结构成功实现了对多个目标波长的相位补偿，各畴区的极化方向排列能够使不同波长的基波在传播过程中始终保持相长干涉，从而保证了有效非线性系数的最大化积累。在计算效率方面，该优化过程在 1000 次迭代中，维度在 3000 时的平均耗时仅为 21 秒。这种效率的显著提升不仅加速了复杂 AOS 结构的设计

进程，更使得大尺寸、多目标的超晶格优化问题具备了实际可操作性，为后续基于该晶体的非线性光学器件集成与应用奠定了重要思想。综合来看，实验结果充分验证了所设计晶体在多波长二次谐波产生场景下的有效性与优越性，其性能指标完全符合预设的设计目标。

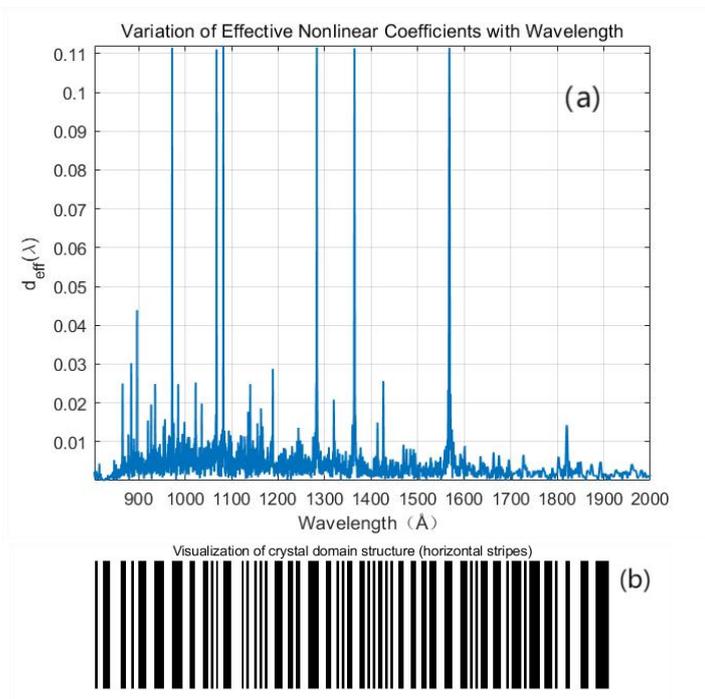

图 7 图 a 展示了设计后的晶体在 972 nm、1082 nm、1283 nm、1364nm 和 1568nm 波长下 $d_{eff}$ 分布图；

图 b 展示了部分晶畴灰度图，最小单元为 3um

在上图中我们还观察到六个顶峰，相当于我们在该晶体中还额外得到一个 1067nm 的 $d_{eff}$，这个结果与前面提到的文献中的结果很相似，上面提到的文献也是同样找到这六个顶峰。在本次设计 PPLN 的实验中平均下来这五个预设的波长的 $d_{eff}$ 平均值在 0.11 左右，偶尔存在一些杂峰在 0.04 左右，大部分波长都在 0.01 以下。

我们还在尝试在在相同晶体长度和相同晶体晶畴下，预设三波长与四波长来验证算法的稳定性，在三波长二倍频(如图 8)下的基波为 972nm、1082nm、1283nm 与四波长二倍频（如图 9）下的基波为 1000nm，1064nm，1082nm 与 1568nm 下的 $d_{eff}$ 分布图以及晶畴灰度图。

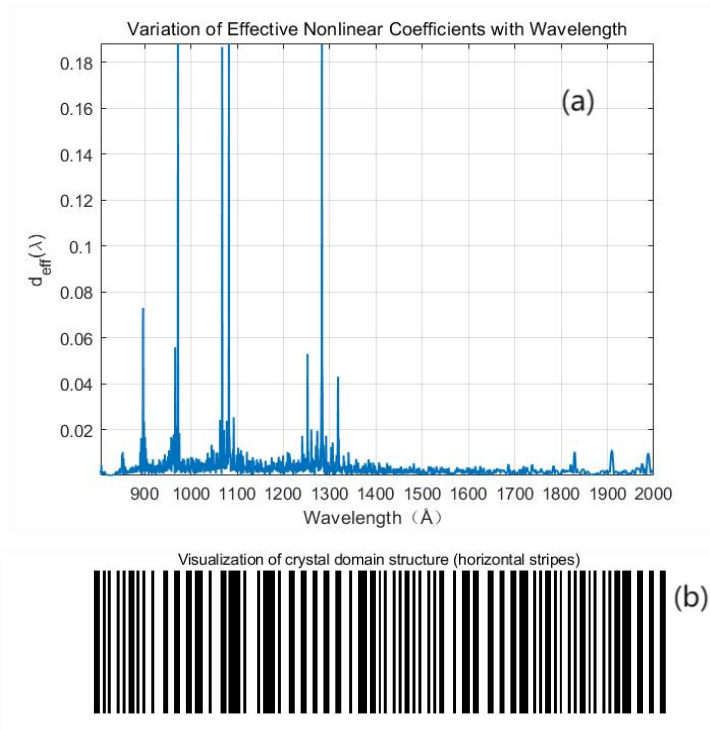

图 8 图 a 展示了设计后的晶体在 972nm、1082nm、1283nm 波长下 $d_{eff}$ 分布图；

图 b 展示了部分晶畴灰度图，最小单元为 3um

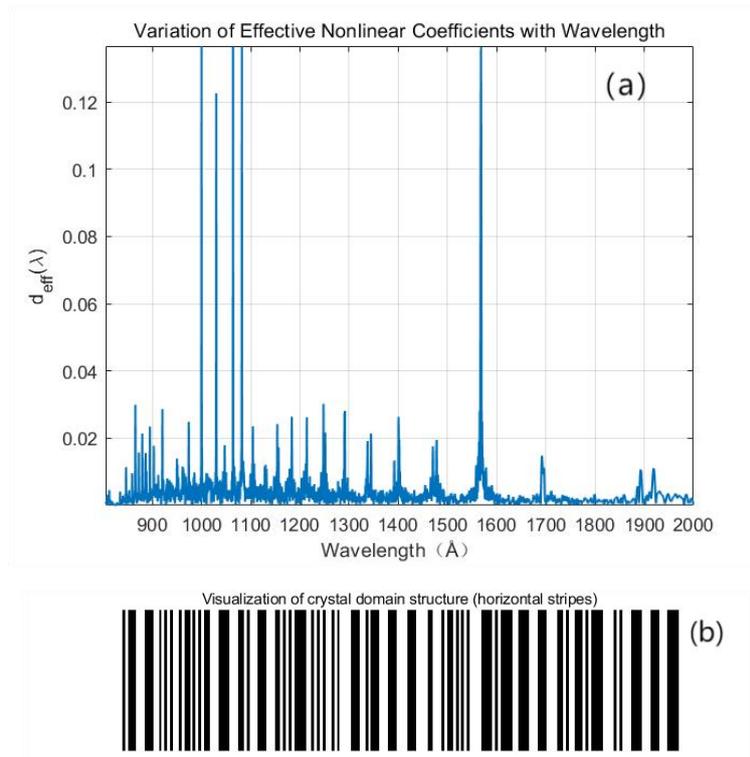

图 9 图 a 展示了设计后的晶体在 1000nm，1064nm，1082nm 与 1568nm 波长下 $d_{eff}$ 分布图；

图 b 展示了部分晶畴灰度图，最小单元为 3um

最后，本研究进一步将所提算法应用于多波长耦合三倍频的准相位匹配晶体设计中。在目标函数构建方面，仍沿用多波长二倍频的形式，仅在有效非线性系

数的求解过程中，采用了本文第二章所推导的特定计算公式。针对晶体长度固定为 9mm 晶畴周期设定为 3μm 的条件，首先对泵浦光波长分别为 1000nm、1300nm 和 1500nm 的三波长耦合场景进行了晶体结构优化设计，并获取了相应的晶体设计结果图及各波长在该晶体中的表现特性图（如图 10 所示）。从图 10 的分析结果可见，所提算法成功搜寻到了能够同时满足这三个波长耦合三倍频需求的最优有效非线性系数配置，且三者的计算结果具有良好的一致性，表明该设计可确保三种波长在同一晶体中均能获得有效的增益输出。此外，在相同的晶体长度（9mm）与晶畴周期（3μm）参数条件下，本研究还开展了泵浦光波长为 1404nm 和 1650nm 的双波长耦合三倍频晶体设计，相关的晶体结构设计结果及性能表征如图 11 所示。

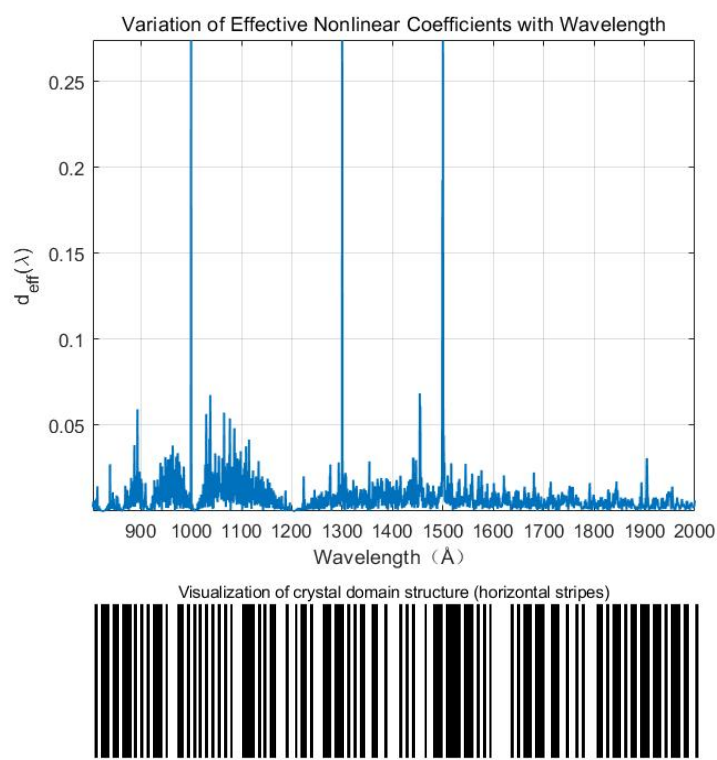

图 10 图 a 展示了设计后的晶体在 1000nm、1300nm 和 1500nm 波长下 $d_{eff}$ 分布图；

图 b 展示了部分晶畴灰度图，最小单元为 3um

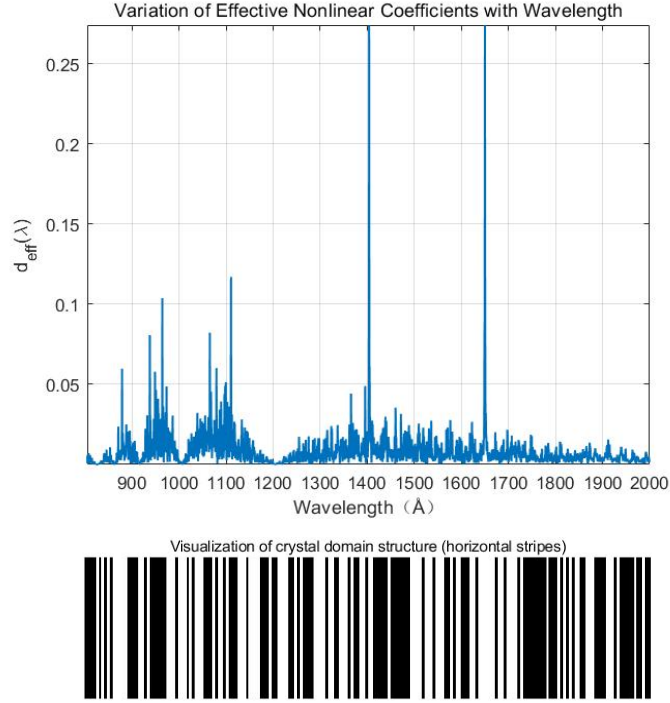

图 11 图 a 展示了设计后的晶体在 1404nm 和 1650nm 波长 $d_{eff}$ 下分布图；

图 b 展示了部分晶畴灰度图，最小单元为 3um

## 6. 结论

CUDA 为差分进化算法提供了理想的并行计算平台，特别是从种群级并行的角度来看，GPU 的大规模并行计算能力、高内存带宽和丰富的优化工具可以充分发挥 DE 算法的潜力。在本文的混合优化算法中，利用 CUDA 可以显著提高算法的执行效率，缩短优化时间，使算法能够处理更复杂、更高维度的实际问题。综上，本研究通过设计 GPU 加速的自适应混合差分灰狼优化算法，旨在解决非周期极化晶体性能优化中的非线性极化晶体设计问题。在解决高维组合优化计算效率瓶颈的同时，实现了算法收敛速度与求解精度的平衡；在保证物理约束严格满足的同时，提升了全局搜索能力与局部开发能力的协同效应。此外，本研究为非线性光学材料的性能调控提供了技术支撑，也为高维组合优化问题的高效求解探索了新路径。

## 7. 参考文献

Press, 1975